\documentclass{iaunew379}

\usepackage{graphicx}
\usepackage{amsmath}
\usepackage{multirow}

\begin{document}

\lefttitle{R.~A.~P. Oliveira et al.}
\righttitle{IAU Symposium 379: Assembly history of the Magellanic Bridge}

\jnlPage{1}{7}
\jnlDoiYr{2023}
\doival{10.1017/xxxxx}

\aopheadtitle{Proceedings of IAU Symposium 379}
\editors{P. Bonifacio,  M.-R. Cioni, F. Hammer, M. Pawlowski, and S. Taibi, eds.}

\title{Ages, metallicities and structure of stellar clusters in the Magellanic Bridge}

\author{Raphael A.~P. Oliveira$^1$, Francisco F.~S. Maia$^2$, Beatriz Barbuy$^1$, Bruno Dias$^3$ \& the VISCACHA collaboration}
\affiliation{$^1$ Universidade de S\~ao Paulo, IAG, Rua do Mat\~ao 1226,
S\~ao Paulo 05508-090, Brazil\\
$^2$ Universidade Federal do Rio de Janeiro, Av. Athos da Silveira, 149,
Rio de Janeiro 21941-909, Brazil\\
$^3$ Instituto de Alta Investigaci\'on, Sede Esmeralda, Universidad de Tarapac\'a, Av. Luis Emilio
Recabarren 2477, Iquique, Chile}

\begin{abstract}
The formation of the Magellanic Bridge
during an encounter between the Magellanic Clouds $\sim 200$\,Myr ago
would
be imprinted in the chemical evolution and kinematics of its stellar population, with sites of active star formation.
Since it contains hundreds of stellar clusters and associations, we combined deep photometry from VISCACHA and SMASH surveys to explore this topic, by
deriving structural parameters, age, metallicity, distance and mass for 33 Bridge clusters with robust statistical tools.
We identified a group of 13 clusters probably stripped from the Small Magellanic Cloud ($0.5-6.8$\,Gyr, $\rm{[Fe/H]}<-0.6$\,dex) and another 15 probably formed in-situ ($< 200$\,Myr, $\rm{[Fe/H]}\sim-0.4$\,dex). Two metallicity dips were detected in the age-metallicity relation, 
coeval to
the Stream and Bridge formation epochs. Cluster masses range from $500$ to $\sim 10^4 M_\odot$, and a new estimate of $3-5\times 10^5 M_\odot$ is obtained for the Bridge stellar mass.
\end{abstract}


\maketitle

\section{Introduction}

The Magellanic System contains the Large and Small Magellanic Clouds (LMC and SMC),
Magellanic Bridge, Stream and Leading Arm.
Besides containing the pair of satellites closest to the Milky Way,
the Bridge is the nearest tidally-stripped structure.
It is also the only one of the three gaseous structures that contain a significant stellar mass \citep[$1.5\times 10^4\,M_\odot$;][]{2007ApJ...658..345H}, with hundreds of star clusters and associations \citep{2020AJ....159...82B}. 
The Bridge was first detected as an HI overdensity
by \citet{1963AuJPh..16..570H},
but a blue, young stellar population was found decades later, more concentrated close to the SMC Wing and halfway the Bridge, and strongly correlated with the distribution and kinematics of the HI gas \citep{2014ApJ...795..108S}. An old population was also found later, more spread across the Bridge.

Based on different epochs of \textit{HST} proper motions, two $N$-body models attempt to reproduce the formation of the LMC-SMC pair and determine if they were independent satellites of the Milky Way until the LMC captured the SMC $\sim 1.2$\,Gyr ago \citep{2011MNRAS.413.2015D}, or an old interacting system in the first perigalactic passage, entering the Galactic potential $\sim 2$\,Gyr ago \citep{2012MNRAS.421.2109B}. Despite being highly dependent on the total masses of the Milky Way and LMC,
both models reproduce the Bridge formation during the most recent encounter between the LMC-SMC \citep[$150-300$\,Myr ago;][]{2012MNRAS.421.2109B, 2018ApJ...864...55Z, 2022ApJ...927..153C}, with gas and stars tidally stripped mostly from the SMC and possibly dragged from the LMC. This scenario would imply a gradient of increasing metallicity toward the LMC due to a minor contribution of its more metal-rich gas, and the presence of an old stellar population amidst a predominant young population formed in situ.

In this work \citep[based on][MNRAS in press]{2023arXiv230605503O}
we analyse deep photometry for 33 Bridge clusters in terms of structural and fundamental parameters in order to investigate their assembly history, spatial distribution and the existence of such gradients.




\section{Photometric data and methodology}

The VISCACHA survey
\citep{2019MNRAS.484.5702M} uses the adaptive optics system of the SOAR 4-m telescope (SAM) to observe clusters in the outskirts of the Magellanic Clouds and Bridge. This very deep ($V\sim 24$\,mag) photometry, with good spatial resolution ($\sim 0.6^{\prime\prime}$), allowed us to derive precise age, metallicity, distance, mass and structural parameters for 33 Bridge clusters at $\rm{RA} < 3^h$ \citep{2023arXiv230605503O}.
Observations with Goodman@SOAR are being carried out to cover the objects in $\rm{RA} > 3^h$. A spectroscopic follow-up in the CaT region was also conducted to derive radial velocities and metallicity for clusters older than $1$\,Gyr (Dias et al., in preparation).
Photometry from the SMASH survey \citep{2017AJ....154..199N} is also used as a complement, with a similar depth but lower spatial resolution in crowded regions.

\begin{figure}
    \includegraphics[width=0.23\textwidth]{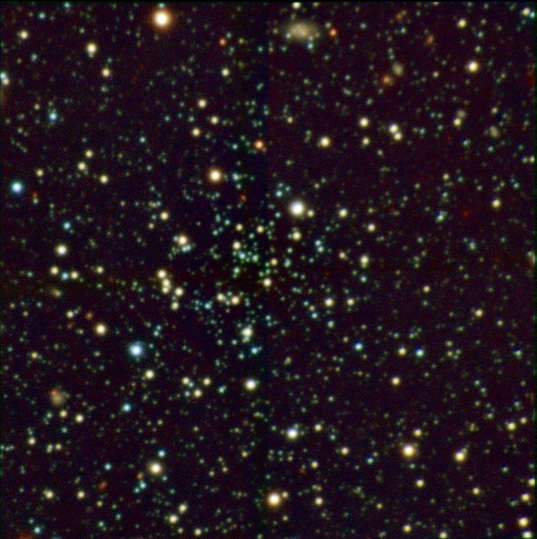}
    \hspace{0.5mm}
    \includegraphics[trim={0 0.2cm 0 0},clip, width=0.435\textwidth]{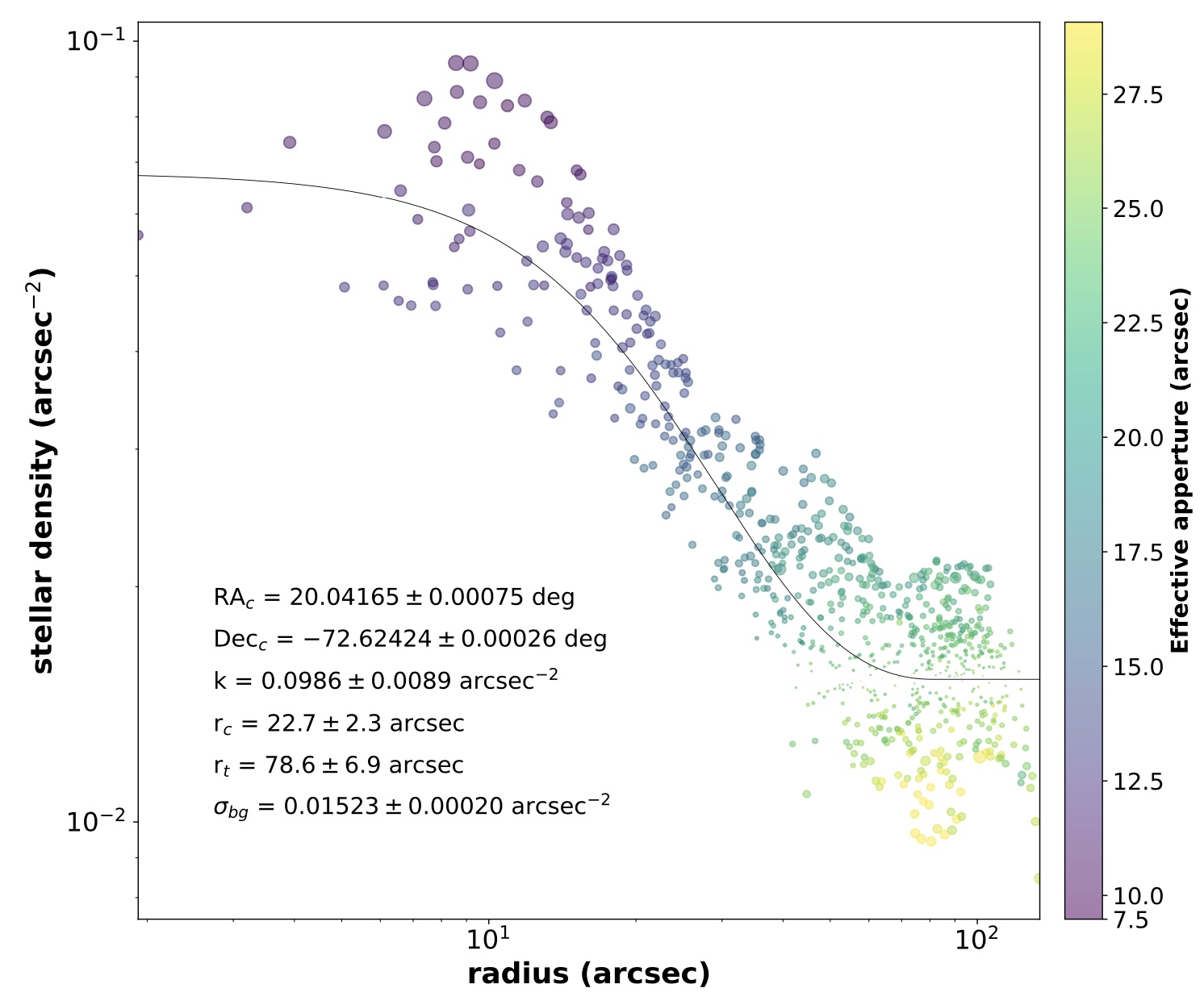}
    \hspace{0.5mm}
    \includegraphics[trim={0 0.4cm 0 0},clip, width=0.297\textwidth]{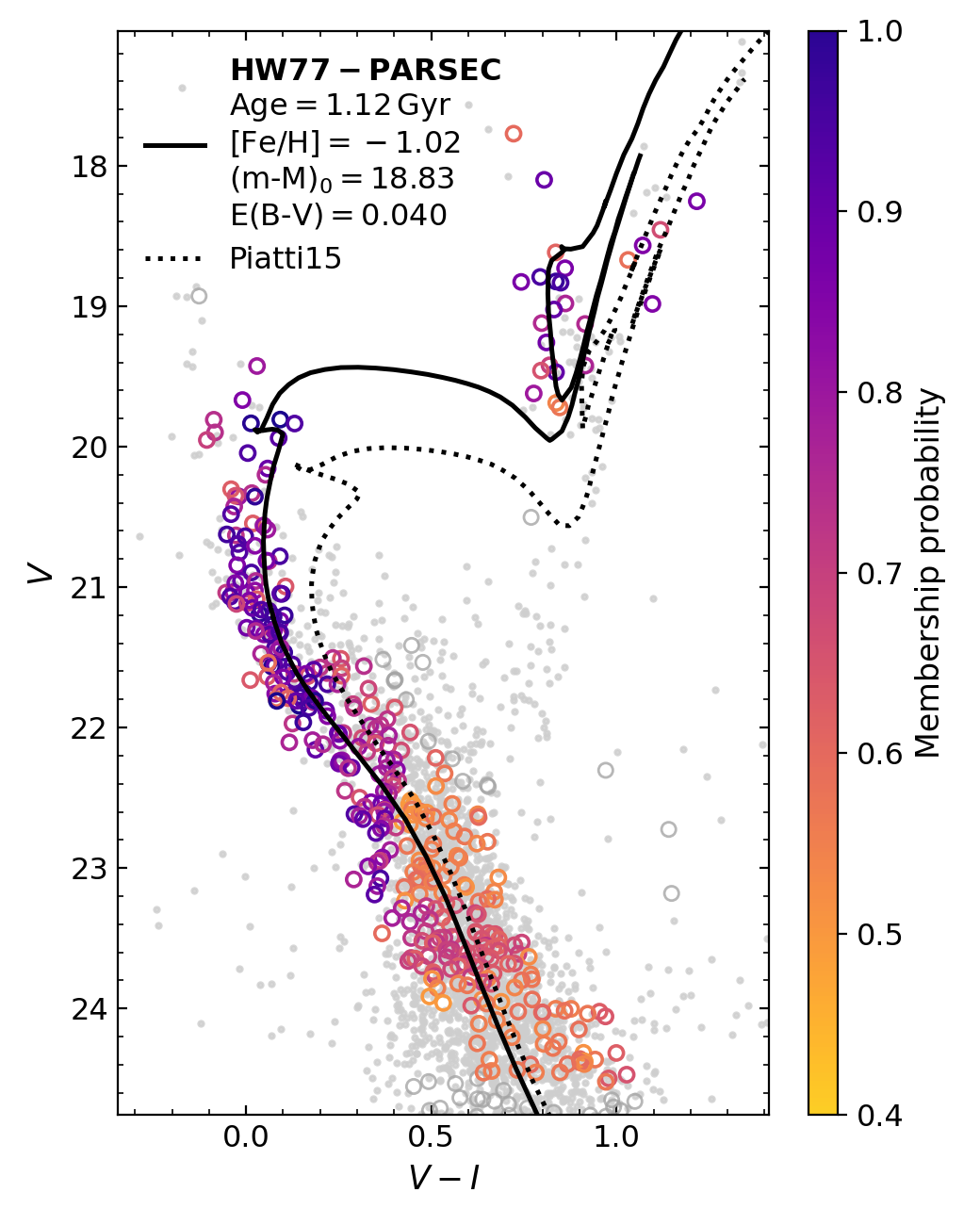}
    \caption{\textit{(Left:)} \textit{BVI} composite image of the SMC Wing cluster HW77, obtained with the SAM@SOAR imager ($3\times 3\,\rm{arcmin}^2$). \textit{(Middle:)} Result obtained in the fitting of King functions to the RDP of HW77. \textit{(Right:)} Isochrone fit obtained for the decontaminated CMD of HW77, with $1.1\pm0.1$\,Gyr and $\rm{[Fe/H]}=-1.02\pm0.11$\,dex, and a final mass of $(2.2\pm0.6)\times 10^3\,M_\odot$.}
    \label{fig:HW77-results}
\end{figure}



In order to ensure a homogeneous and self-consistent analysis, we implemented the Markov chain Monte Carlo (MCMC) sampling \citep{2013PASP..125..306F} both to the fitting of analytical functions to radial density profiles (RDPs) and statistical isochrone fitting to colour-magnitude diagrams (CMDs).
The RDPs
are obtained by computing the local stellar density around each star, using a variable aperture size. A likelihood function comparing the observed local density with the model \citep{1962AJ.....67..471K} is evaluated and coupled to the MCMC to obtain a new centre, core and tidal radii, and central and background densities. Before the isochrone fitting, a membership analysis to exclude probable field stars are performed as described in \citet{2010MNRAS.407.1875M}.
The method compares the stars within a fraction of the tidal radius (cluster + field stars) with a nearby field and, based on their relative location in the CMD and radius to the cluster centre,
computes a median membership value. For the isochrone fitting, we employ the \texttt{SIRIUS} code \citep{2020ApJ...890...38S} to fit PARSEC isochrones \citep{2012MNRAS.427..127B} to the decontaminated
$V$ vs. $V-I$
CMDs. In this case, the likelihood
compares the position of each star in the CMD to the closest point of the tentative isochrone, so that the MCMC retrieves posterior distributions in age, metallicity, distance and reddening.
The cluster mass is obtained by integrating the flux of all member stars, converting it to absolute magnitude and applying a calibration with age and metallicity provided in \citet{2014MNRAS.437.2005M}. 







\section{Results: cluster parameters, gradients and age-metallicity relation}

Figure~\ref{fig:HW77-results} gives the structural and fundamental parameters derived for the Wing/Bridge cluster HW77 with VISCACHA data. The observation was taken in optimal conditions for the use of adaptive optics, providing the deepest CMD and best seeing ($\sim0.5^{\prime\prime}$) of the sample. A large core radius of $22.7^{\prime\prime}$ was derived with a small concentration parameter $c=\log(r_t/r_c) = 0.54$, whereas the age of $1.12$\,Gyr and $\rm{[Fe/H]}=-1.02$\,dex resulted in a mass of $2.2\pm0.6 \times 10^3\,M_\odot$.
The SMASH data for this cluster are nearly identical to VISCACHA, with the CMD $g$ vs. $g-i$ with very similar results and uncertainties (Oliveira et al., in preparation).

The sample of 33 Wing/Bridge clusters resulted in concentration parameters between $0.46$ and $1.25$ (smaller than Galactic open and globular clusters, but consistent with the Magellanic Clouds clusters), ages from $4$\,Myr (HW81) to $6.8$\,Gyr (HW59), distances from $50$ to $69$\,kpc, and masses from $\sim 500$ (L101, ICA45) to $1.4\times 10^4\,M_\odot$ (L113, L110).
The mass of the 33 clusters add up to $10^5\,M_\odot$ which, extrapolating to the $\sim 100$ clusters and $\sim 300$ associations located at the Bridge, provide a new estimate of $3-5 \times 10^5\,M_\odot$ for the Bridge stellar mass, more than one order of magnitude higher than \citet{2007ApJ...658..345H}.

\begin{figure}
    \centering
    \includegraphics[height=4.77cm]{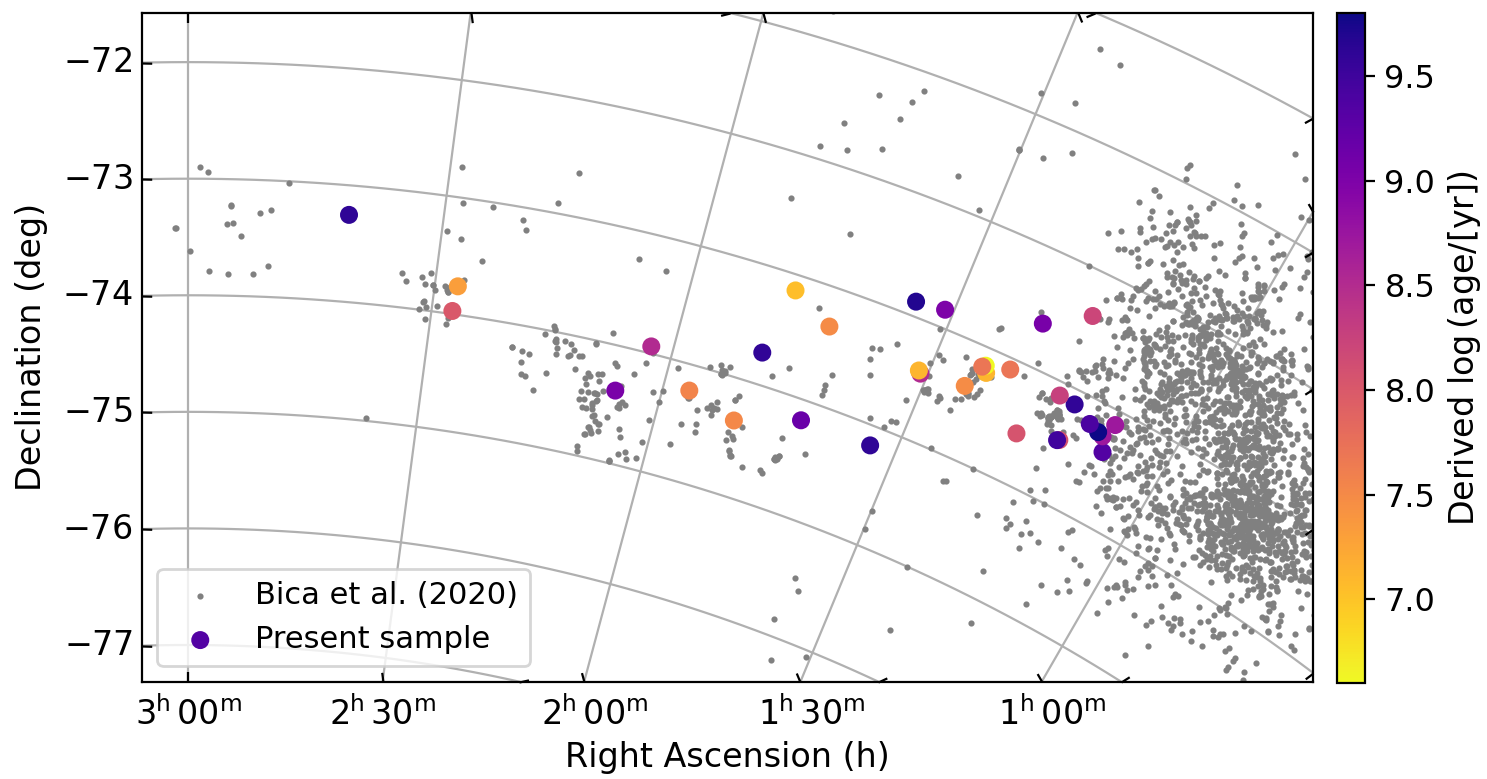}
    \includegraphics[height=4.77cm]{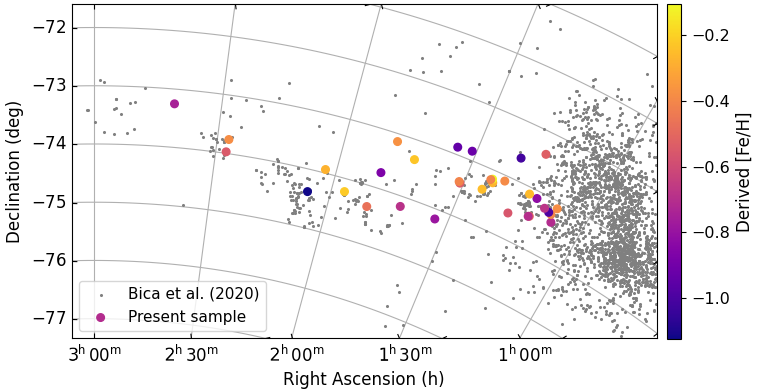}
    \caption{Projected distribution of the derived ages and metallicities for the 33 sample clusters.}
    \label{fig:agesRADEC}
\end{figure}

\begin{figure}
    \centering
    \includegraphics[width=0.60\textwidth]{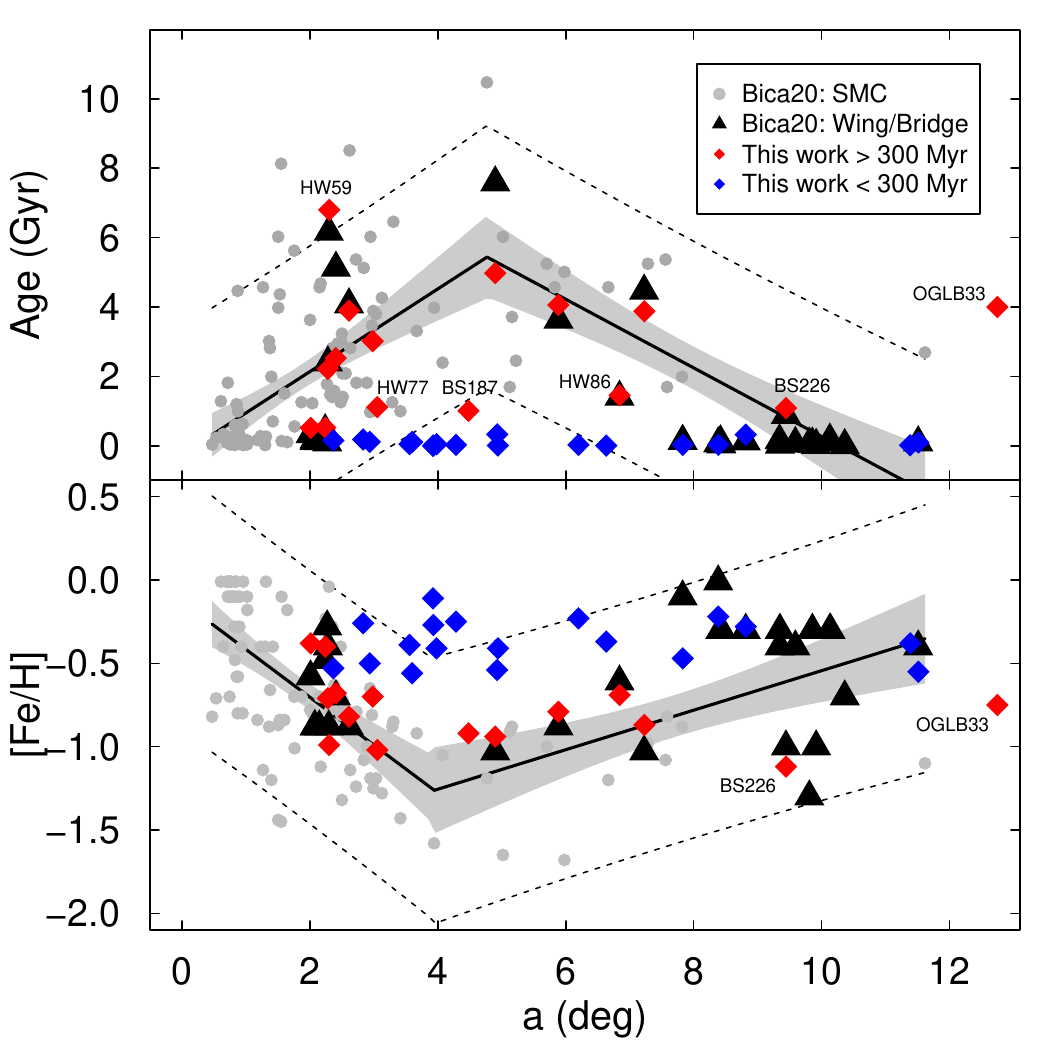}
    \caption{Age and metallicity as a function of the projected distance to the SMC centre. Grey and black points, as well as the lines and shaded area (gradients), are from \citet{2020AJ....159...82B}, and the coloured points are the present sample. The clusters older than $300$\,Myr that deviate from the gradients are identified.}
    \label{fig:grads}
\end{figure}

Figure~\ref{fig:agesRADEC} presents the projected distribution of all the objects from \citet{2020AJ....159...82B} together with the 33 sample clusters colour-coded by the derived age and metallicity. The older clusters are mostly isolated, located close to the SMC, whereas the young ones appear to be grouped along the Bridge. Two groups become evident: old clusters ($>500$\,Myr) more metal-poor than $-0.6$\,dex, vs.  clusters younger than the Bridge, with $\rm{[Fe/H]}\sim -0.4$\,dex, probably formed in situ. Figure~\ref{fig:grads} reproduces a figure from \citet[gray and black dots]{2020AJ....159...82B} with the sample clusters older or younger than $300$\,Myr as red or blue symbols, in order to check if our results follow the age and metallicity gradients of the SMC vicinity. As expected, the old clusters (probably stripped from the SMC) follow both gradients, with increasing age and decreasing metallicity until $a\sim4^\circ$ and an inversion after that. The young clusters does not show a pattern in age and have a nearly constant metallicity along the Bridge.

The age-metallicity relation (AMR) is a valuable tool to analyse the chemical evolution of a galaxy, with hints of chemical enrichment or decrease in metallicity. The present results follow the chemical evolutionary models in most of the cases, but some clusters imprint two metallicity dips in the AMR: a larger one around $1.5$\,Gyr with the metallicity decreasing $0.4$\,dex, and a smaller one around $200$\,Myr ago, with a $0.3$\,dex decrease in metallicity. These epochs coincide with those of the formation of the Magellanic Stream and Bridge, respectively. According to the models \citep[e.g.][]{2009ApJ...700L..69T}, such metallicity decrease is usually explained by an infall of more metal-poor gas, followed by a rapid chemical enrichment (and possibly an increase in the star formation rate).

A complete interpretation of the results is given in \citet{2023arXiv230605503O}.

%





\end{document}